\def\year{2021}\relax
\documentclass[letterpaper]{article} 
\usepackage{multirow}
\usepackage{aaai21}  
\usepackage{times}  
\usepackage{helvet} 
\usepackage{courier}  
\usepackage[hyphens]{url}  
\usepackage{graphicx} 
\def\UrlFont{\rm}  
\usepackage{natbib}  
\usepackage{caption} 
\usepackage[skip=2pt]{caption}

\usepackage{amsmath}
\usepackage{scalerel}
\usepackage{xcolor}
\usepackage{booktabs}
\usepackage{longtable}
\definecolor{nonpol}{rgb}{0.772, 0.353, 0.067}
\definecolor{pol}{rgb}{0, 0.44, 0.759}

\title{
Political Discussion is Abundant in Non-political Subreddits (and Less Toxic)}
\author{Ashwin Rajadesingan,
Ceren Budak,
Paul Resnick\\}

\affiliations{
School of Information\\
University of Michigan, Ann Arbor\\
arajades,cbudak,presnick@umich.edu

}

\begin{document}
\maketitle

\begin{abstract}
Research on online political communication has primarily focused on content in explicitly political spaces. In this work, we set out to determine the amount of political talk missed using this approach. Focusing on Reddit, we estimate that nearly half of all political talk takes place in subreddits that host political content less than 25\% of the time. In other words, cumulatively, political talk in non-political spaces is abundant. We further examine the nature of political talk and show that political conversations are less toxic in non-political subreddits. Indeed, the average toxicity of political comments replying to a out-partisan in non-political subreddits is even less than the toxicity of co-partisan replies in explicitly political subreddits.

\end{abstract}

\section{Introduction}
Casual everyday political conversations are central to a vibrant deliberative democracy. Through these conversations, individuals learn new perspectives, form informed opinions and update their preferences \cite{kim2008theorizing}. These interactions may take place in explicitly political spaces such as city townhalls and civic committees but also in seemingly non-political spaces such as book readings, workplaces and social gatherings \cite{conover2018taking}. Importantly, this kind of everyday political talk is significantly correlated with opinion quality and political participation which are central to forming a well-informed electorate \cite{wyatt2000bridging}. In this work, we explore this phenomenon online, particularly studying political discussions in communities on Reddit that are not explicitly political. 

Most research on political discussions has primarily focused on explicitly political spaces, examining communities around political news groups, figures or ideologies \cite{soliman2019characterization,himelboim2009discussion,an2019political}. However, survey research suggests that most people encounter political content online not in explicitly political spaces but in hobby and leisure groups where politics is incidental to the conversation \cite{wojcieszak2009online}. Further, recent years have seen increased political engagement among the electorate perhaps due to high levels of partisanship \cite{huddy2015expressive} and growing social movements \cite{bosi2016consequences} such as Black Lives Matter. This heightened level of political engagement can also be observed online. For example, on Reddit, many communities that would not be typically construed as being `political' such as r/EDM and r/MaleFashionAdvice protested against the platform’s hate speech policies and police brutality in the US.\footnote{https://www.theverge.com/2020/6/3/21279601/reddit-dark- subreddits-protest-police-violence-racism-hate-speech-policies} Further, in recent years, scholars have observed increasing politicization of typically non-political spaces \cite{dagnes2019us}. The most prominent example is the politicization of emerging science and technology where inherent uncertainties are harnessed by political actors to cast doubt on the existence of scientific consensus \cite{bolsen2015counteracting}. On Reddit, this phenomenon manifests in the formation of multiple communities on the same topic along partisan lines, for example, r/China\_Flu and r/Coronavirus \cite{zhang2020tale}. Thus, these developments call for expanding analysis of political discussions outside of typical political communities to communities that aren't explicitly political.

It is important to note that expanding the study of political discussions to include non-political spaces does not merely increase the volume of discussions for analysis. The dynamics of political discussions in these spaces may also be fundamentally different. Political discussions in these spaces may be moderated not by partisan identity but by participants' shared non-political interests and identities that drew them to the same community in the first place \cite{gaertner2011common}. Thus, we might expect political conversations in non-political spaces, including cross-partisan ones, to be less toxic. 
However, 
shared non-political group identity may fail to offset, and might even exacerbate, the animosity generated by partisan identity \cite{klar2018common}. Further, norms in these non-political spaces may not be designed to foster political discourse. Indeed, there may be norms against having political conversations at all, and thus when they occur, they may be even more toxic.

In this work, we focus on two primary questions: (i) What is the prevalence of political discussions in communities that are not explicitly political? (ii) Are cross-partisan political discussions in these spaces less toxic than ones in explicitly political spaces? We estimate that 49.26\% $\pm 3.59$\%  of all political discussions on Reddit takes place in communities that host political discussions less than 25\% of the time. This finding is not simply the result of a few very large non-political communities hosting some political content. It is instead due to a long tail of small communities that host some political content each. 
Our toxicity analysis reveals that political conversations in non-political spaces, including cross-partisan political interactions, are indeed less toxic than such interactions in political spaces. Interestingly, we find that there is an uptick in toxicity levels when talking politics, but even with this increase, the toxicity levels in non-political subreddits are still much lower than the toxicity in political subreddits.

\section{Background}

Political scientists have long highlighted the presence and importance of casual political talk in everyday social interactions taking place in spaces that are not explicitly political (see \cite{conover2018taking} for a review). In fact, research suggests that most political conversations take place at work or with neighbors, with more than 70\% of American survey respondents reporting that they have never or only rarely even attended public meetings explicitly designed for political discussions \cite{conover2002deliberative}. Similarly online, early survey research suggests that most people encounter political talk in message boards and chatrooms designed not for political discussions but for hobby and leisure related discussions \cite{wojcieszak2009online}. Thus, research limited to studying only political discussion spaces may overlook other spaces where a significant amount of such interactions may be taking place. Such everyday political talk, although not always deliberative and conducive to rational-critical argumentation, have important positive outcomes such as increased political knowledge \cite{pattie2008s}, political participation \cite{searing2007public}, refined opinions \cite{kim2008theorizing} and higher tolerance \cite{pattie2008s,mutz2006hearing}. 

Recently, scholars have analyzed political discussions taking place in online ``third spaces'' a term derived from sociologist Ray Oldenburg's conceptualization of the 'third place', referring to public spaces outside of work and home such as cafes, parks and libraries where people meet and interact informally, fostering community ties and political participation \cite{wright2012third,oldenburg2001celebrating}. Graham et al. \citeyearpar{graham2015everyday} found that political discussions in the three UK-based non-political forums they analyzed were as likely to emerge from non-political, personally-oriented discussions as from discussions that were about politics from the start, with users explicitly linking their personal experiences to public policy. In contrast to discussions in political spaces, they found that the discursive culture in these discussions centers around help and support rather than being competitive and combative. 
Yan et al. \citeyearpar{yan2018s} found that the political arguments made on transnational online cricket forums were typically short, unsubstantiated by external sources and occasionally uncivil. However, there was high exposure to cross-cutting political discussions with some engagement with opposing views in the form of question exchange and mutual acknowledgement. Analyzing a  reality television discussion forum, Graham found that most political exchanges were driven by users' life experiences representing a more ``lifestyle-oriented, personal form of politics'' \cite{graham2012beyond}. While exhibiting deliberative features such as the exchange of reasoned claims (as opposed to assertions) and reciprocity, participants also employed affirming, supportive and empathetic communicative practices fostering genuineness and civility in the discussions.

Political discussions in these non-political spaces may also be more civil and social compared to discussions in explicitly political communities. A significant factor contributing to hostility commonly observed in online political discussions is the increased levels of affective polarization \cite{hutchens2019reinforcing}, the tendency of partisans to view opposing partisans negatively and co-partisans positively \cite{iyengar2015fear}. This increased out-party animosity is explained by Social Identity Theory which argues that by merely categorizing individuals into groups (here, Republicans and Democrats), group identities are activated, creating an ‘us’ versus ‘them’ group dynamic \cite{tajfel1979integrative}. Crucially, unlike race, gender and other protected attributes where group-related behaviors are mediated by strong social norms (and laws) against discrimination, there are no norms that temper hostility towards out-partisans \cite{iyengar2019origins}. In fact, the open hostility displayed by political elites towards their political opponents demonstrates that such behavior is appropriate \cite{banda2018elite}. Given the social identity underpinnings of affective polarization, researchers have explored ways to offset partisan identity drawing from prior research on intergroup conflict. One successful approach to reducing out-partisan animosity is by priming a superordinate identity. Based on the Common Ingroup Identity Model \cite{gaertner2011common}, Levendusky showed that priming Republicans and Democrats to think of each other as Americans rather than outgroup members recategorized them as being part of the same common ingroup, resulting in reduced animosity and warmer attitudes towards each other \cite{levendusky2018americans}.  Although our study is not a direct test of this theory, we expect that interactions in non-political subreddits likely increase the salience of shared common non-political group memberships. This may mediate how cross-partisan political discussions are conducted in these spaces.

Though not conclusive, the prior literature provides two compelling arguments: (i) political conversations are abundant in non-political spaces. (ii) quality of discourse in these conversations may be different and in some cases, better than political conversations in explicitly political communities. In this work, we assess these claims empirically in the context of Reddit. First, we quantify the relative contribution of non-political subreddits to the overall political content on Reddit. In this aspect, our work is similar to Munson et al.'s work on estimating the prevalence of political content in non-political blogs. Analyzing a sample of blogs from Blogger.com, they found that ``25\% of all political posts are from blogs that post about politics less than 20\% of the time'' \cite{munson2011prevalence}. Second, we examine a specific marker of conversation quality: toxicity in cross-partisan political interactions. Scholars have suggested that political talk in these third spaces are likely to be less polarized, since users participate in these spaces because of shared interests such as a soccer team or fast fashion which are not aligned politically \cite{wojcieszak2009online,wright2015third}. Thus, mediated by shared non-political identities, these spaces could facilitate respectful and civil cross-partisan interactions. In this work, we examine this hypothesis by quantifying the toxicity levels of cross-partisan political discussions in non-political spaces and comparing them to toxicity levels in other settings on Reddit.

\section{Reddit Data}
Reddit, a collection of communities of varied and diverse topics, provides us with an ideal platform to examine the prevalence of political content in non-political spaces. We use the PushShift Reddit dataset \cite{baumgartner2020pushshift} to perform our analysis on comments posted from 2016 to 2019. We exclude comments from subreddits that have hosted less than 1000 comments over the four years. We also remove comments from known bots and moderators from the analysis. To allow for robust estimation of political prevalence, we only consider comments which are 50 characters or more in length in this analysis. In total, we examined 2.8 billion comments posted in 30,899 subreddits.

\section{Estimating the Prevalence of Political Content in Non-political Spaces}\label{framework}

Our basic approach to estimating the prevalence of political content is to train a classifier that yields, for each comment, a predicted probability that it would be judged as political by a panel of three MTurk raters. If the classifier's outputs are properly calibrated, the average of those outputs for all the comments in a subreddit is an estimate of the prevalence of political content in the subreddit.

Our training data consists of a sample of 10,000 comments, each rated by three people on MTurk as either political or not. 
We do not use these labels to directly train a classifier that predicts what how each item will be labeled. Following the quantification approach \cite{forman2006quantifying,gonzalez2017review}, if the goal is to estimate prevalence rather than to correctly classify individual comments, it can be more effective to use ground-truth data to perform calibration on a crudely trained classifier than to use up the training data on improving the classifier.

\begin{figure*}[!ht]
\includegraphics[scale=0.55]{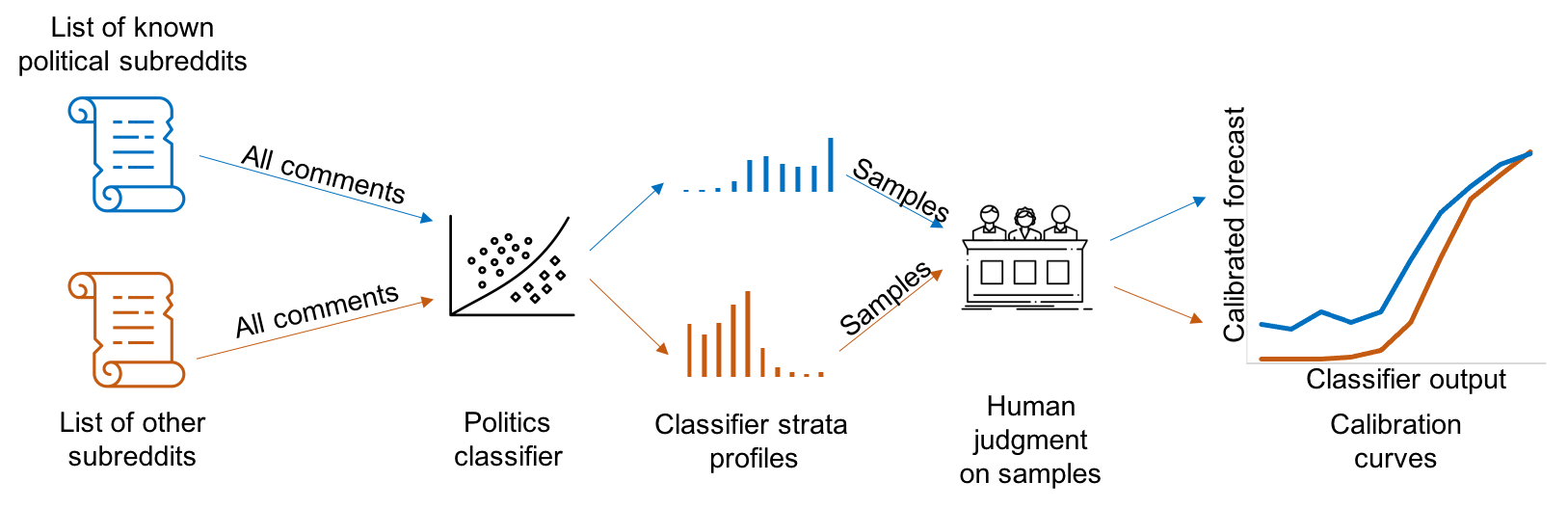} 
\centering
\caption{Training a classifier that distinguishes between comments from political and non-political subreddits, then calibrating it to produce predictions of whether comments are political.}
\label{fig:building_calibrators}
\end{figure*}

Section~\ref{training_classifier} describes a process for training a classifier to distinguish between comments from two baskets of subreddits, one of which consists of a hand-selected set of subreddits that are overtly political. This text-based classifier outputs a probability that the comment is from one of the political subreddits. The middle of Figure~\ref{fig:building_calibrators} shows the distribution of classifier outputs for all comments from the two baskets of subreddits. Some comments from the political subreddits contain phrases that are more common in the other subreddits. Such comments get a low score from the classifier. However, most comments originating in the political subreddits get higher scores (the blue distribution, on top) and most comments originating in the other subreddits get lower scores (the orange distribution, below).

The classifier does not have perfect accuracy. Moreover, it may make different kinds of errors on content from different subreddits. So we do not directly use it to classify and count political comments in each subreddit. The second step, as described in section~\ref{building_calibrators}, is to build two calibration curves for the classifier, one based on human ratings of a sample of comments from the set of known political subreddits and the other based on human ratings for comments from other subreddits. Each calibrator provides a mapping from a classifier output stratum (e.g., 0.5-0.6) to a calibrated forecast, the fraction of comments that are political when the classifier gives an output in that stratum. The right side of Figure~\ref{fig:building_calibrators} shows the two calibration curves. For all classifier outputs below 0.9, comments originating in the political subreddits were more likely to be judged as political, as indicated by the gap between the blue calibration curve and the orange one.

Section~\ref{selectcalib} then describes a process for selecting a calibrator to use for a particular subreddit. Section~\ref{estimate} describes how to use the calibrator to generate an estimate of the fraction of political content in that subreddit. Finally, section~\ref{cumulative} describes how we combine the estimates for individual subreddits to yield overall prevalence.

\subsection{Training a Classifier}\label{training_classifier}

We built an L1-regularized logistic regression classifier trained on bigrams and trigrams from a random sample of 500,000 comments from known political subreddits (positive or ``politics'' class) and 500,000 comments from all other subreddits (negative or ``not politics'' class). We used a list of 277 political subreddits provided by \cite{rajadesingan2020quick} and updated the list to include more recently created subreddits supporting Democratic primary candidates such as r/YangForPresidentHQ and r/JoeBiden before the 2020 US presidential election. 

Assessed through 5-fold cross-validation, we obtained an accuracy of 81.56\% with a false positive rate of 14.41\% and a false negative rate of 22.45\%. Note that the false positive and false negative rates are for predicting the source of a comment, not whether the comment itself is truly political. While this classifier performed reasonably well in identifying content from political subreddits, it was trained not on political and non-political comments but on \textit{comments from political and non-political subreddits}.
This is particularly problematic since our goal is to estimate the fraction of \textit{political comments} in non-political subreddits. If the classifier were perfectly accurate at distinguishing between content from the two types of subreddits, and we used it as if it were distinguishing political comments, it would tell us that there were zero political comments in non-political subreddits, which might or might not be the case. Thus, a further calibration step is needed in order to estimate the error rates of this classifier at predicting whether a comment is truly political, and then adjust for those error rates in our prevalence estimates.

\subsection{Building Calibrators}\label{building_calibrators}

\begin{table*}[t]
\label{eval}
  \centering
  \begin{tabular}{|c|c|c|c|c|c|c|c|c|}
    \hline
    \multirow{3}{*}{\textbf{Strata}} & \multicolumn{4}{c|}{\textbf{Political subreddits}} &
    \multicolumn{4}{c|}{\textbf{Non-political subreddits}}\\
        \cline{2-9}
    & \textbf{Pop. proportion} & \textbf{Samples} & \textbf{Political\%} & \textbf{Std dev.} &\textbf{Pop. proportion} & \textbf{Samples} &\textbf{Political\%}& \textbf{Std dev.}\\
   & $W_{k,\textrm{pol}}$&   & $p_{k,\textrm{pol}}$ & $s_{k,pol}$ &$W_{k,\textrm{nonpol}}$ &  & $p_{k,\textrm{nonpol}}$& $s_{k,nonpol}$\\
    \hline
 1 & 0.004 & 50 & 0.180 & 0.054 & 0.148 & 615 &   0.021 &  0.006\\
 2 & 0.007 & 50 & 0.160 & 0.052 & 0.117  & 797   & 0.024 &  0.005\\
 3 & 0.017 & 50 & 0.240 & 0.060 & 0.150 & 1239 &  0.023 &  0.004\\
 4 & 0.047 & 107 & 0.187 & 0.038 & 0.199 & 1810 &  0.031 &  0.004\\
 5 & 0.145 & 346 & 0.237 & 0.023 & 0.242 & 2296 & 0.057 &  0.005\\
 6 & 0.165 & 394 & 0.475 & 0.025 & 0.083  & 788   & 0.189 &  0.014\\
 7 & 0.129 & 295 & 0.695 & 0.027 & 0.026  & 237 & 0.485 &  0.032\\
 8 & 0.116 & 241 & 0.821 & 0.025 & 0.013  & 107 & 0.757 &  0.041\\
 9 & 0.119 & 204 & 0.917 & 0.019 & 0.008  & 61 & 0.869 &  0.043\\
 10 & 0.252 & 263 & 0.970 & 0.011 & 0.012  & 50 & 0.980 &  0.020\\
    \hline
  \end{tabular}
  \caption{Population proportions, Neyman samples, percent of samples identified as political by human judgment and standard deviation per strata for political and non-political subreddits. The standard deviations for both political and non-political subreddits are lower in the strata where their population proportions are higher. This effect is by design (using Neyman allocation) to ensure that the confidence intervals for the prevalence estimates are lower. }\label{tab:sampling_estimation}
\end{table*}
The right side of Figure \ref{fig:building_calibrators} outlines the calibration process. We use the classifier to produce a probability estimate for each comment. Then, we allocate comments into ten strata, 0-10\%, 10-20\%, etc., based on the classifier outputs. That yields a profile of classifier strata: the proportion of comments that fall into each stratum. We compute two separate classifier strata profiles, one based on comments from the list of known political subreddits, the other based on comments from other subreddits. As might be expected, the classifier assigns many more comments from the political subreddits to the higher strata (higher probability of being political).

Then, we calibrate the classifier outputs against human judgments of the comments, separately for comments from each source. Below we first describe the human judgment process and then explain the rationale and details behind each of the steps in the calibration process.

\subsubsection{Human judgments}
When asked to identify topics they considered political from a list, Fitzgerald \citeyearpar{fitzgerald2013does} found systematic demographic differences with partisans, liberals and men identifying significantly more topics as political compared to non-partisans, conservatives and women respectively. In order to reduce such differences in interpretation, Fitzgerald suggests providing human raters with an explicit definition to follow.

For this work, we modify \cite{moy2006predicting}'s political discussion definition, which is predominantly based on political issues, to also include references to political figures, parties and institutions. We consider a comment to be political if it is about (i) political figures, parties and institutions, (ii) Broad cultural and social issues (e.g., civil rights, moral values, and the environment), (iii) National issues (e.g., healthcare, welfare policy, and foreign affairs), (iv) Local and state concerns (e.g., school board disputes and sales taxes) or (v) neighborhood and community affairs (e.g., decisions about a neighborhood watch crime prevention program). 

Even with an explicit definition, however, whether a particular comment is political or not is open to interpretation. Conceptually, we take the ground truth classification of a comment to be the label that the majority of people who ever read online comments would apply to that comment, if they all were asked to judge it according to the explicit definition.
Of course, this ground truth is a counterfactual; no such survey of all readers of comments can ever be conducted for any comment. Instead, we rely on a proxy for this ground truth, a survey of three raters on Amazon Mechanical Turk. To elicit high quality labels, we limit the task to crowdworkers with high performance in prior tasks \footnote{Raters must be based in the US, previously completed 1000 tasks and have at least 98\% acceptance rate on the tasks that they have previously completed.} who also correctly labeled at least 4 out of 5 items in a qualification task where raters were shown sample comments and were asked to identify if the comment was political or not. Such qualification tasks are shown to improve crowdsourcing label quality~\cite{budak2016fair}.

The inter-rater agreement score as computed by Krippendorff's alpha was 0.55. While an alpha score of 0.55 is just below the threshold used for conventional content analysis, this agreement is relatively higher compared to other cases of crowd coding (e.g.~\cite{lind2017content}). The most common outcome (66.85\%) was for a comment to be unanimously labeled as not political. An additional 10.81\% of comments were unanimously labeled as political. The remainder were split decisions: 7.75\% were labels as political by two of three raters and 14.59\% by one of three raters. 

Following common practice in treatment of crowd labels, our primary analysis treats a comment as truly political if two or three of the raters label it as such. Given the relatively low agreement among raters, for robustness, we also report analyses in the Appendix that treat a comment as political if any of the three label it as political, or only if all three label it as political. The different aggregation strategies produced largely similar results and provide additional informative bounds on our estimates.

\subsubsection{Classifier strata}

We group comments into ten strata based on the classifier probability output. For example, stratum 1 consists of all comments whose classifier output is between 0 and 0.1; stratum 10 consists of all comments whose classifier output is between 0.9 and 1. By stratifying comments into multiple relatively homogeneous groups based on classifier probability, we require fewer samples to estimate true prevalence per stratum as within-group variance reduces in more homogeneous groups \cite{Cochran_1977}. 

\subsubsection{One calibration for each subreddit type}

We expect the per-stratum prevalence estimates to be different in different subreddits. That is, comments in the 60-70\% stratum in political subreddits could be judged 70\% of time to be political subreddits, while this number can be, say, 45\% in non-political ones.  
This would not be a concern if we were trying to estimate the overall prevalence of political comments, since we could just estimate the classifier's error rates on a random sample of comments. However, our task demands accurate estimates of political prevalence in each subreddit; if the classifier is more prone to err on content from the stratum that originates in one subreddit than another, it would throw off our cumulative prevalence estimate. 

Estimating separate per-stratum prevalence rates for each subreddit is practically infeasible as it would require human judgments for samples from each subreddit. Instead, we compute per-stratum error rates separately for each \emph{subreddit type}: a sample of comments from known political subreddits and a sample of comments from other subreddits (same as those used to train the classifier). 

\begin{figure*}[ht]
\includegraphics[scale=0.6]{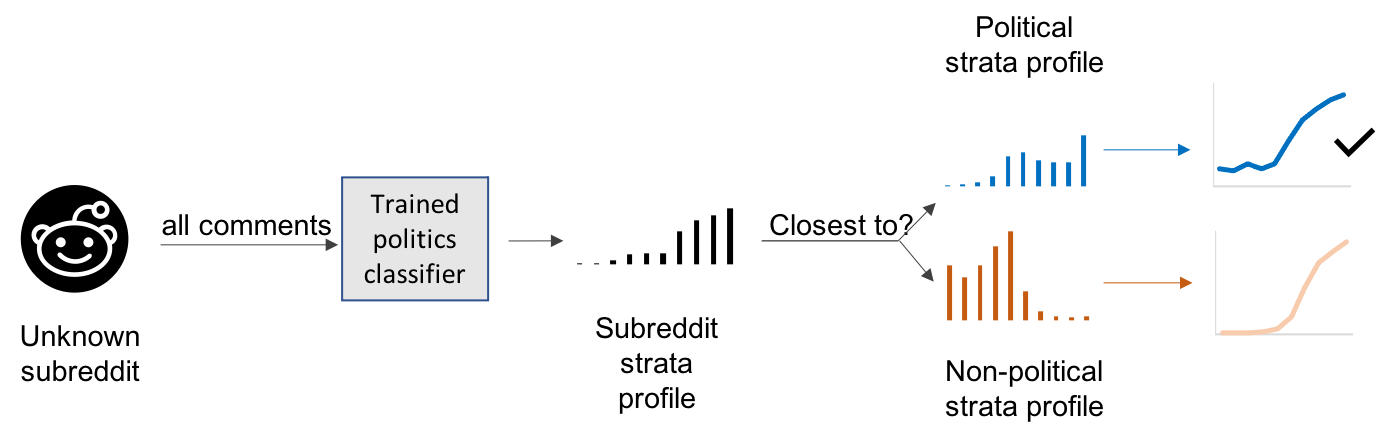} 
\centering
\caption{Selecting the political or nonpoltical calibrator depending on if the subreddit strata profile is similar to the political or non-political strata profile}
\label{fig:selecting_calibration}
\end{figure*}

\subsubsection{Optimum strata sampling for human judgments}
For each subreddit type, we must sample comments from each stratum for human judgments to obtain stratum specific prevalence estimates. The more comments we sample from a stratum, the less variance there will be in our estimate of the true prevalence of political comments in that stratum. 
Intuitively, fewer samples should be taken from a stratum with very few comments for calibration. A high variance in our estimated prevalence of political comments in such a stratum will not affect our overall estimate very much because it affects very few comments.
Formally, for a fixed number of comments that we can afford to send for human rating, Neyman allocation provides the optimal allocation strategy which minimizes the variance of the overall prevalence estimate. Under Neyman allocation, the number of samples allocated to each stratum is given by:
 \begin{displaymath}
n_k = n\frac{W_k * S_k}{\sum_{i=1}^{K}W_i * S_i}
\end{displaymath}
\begin{itemize}
    \item $n$ is the total number of comments to be rated
    \item $K$ is the number of strata (10 in our case)
    \item $n_k$ is the number of comments to sample from the $k$-th stratum
    \item $W_k$ is the weight of the $k$-th stratum in the classifier strata profile, i.e. the fraction of comments that are in that stratum
    \item $S_k$ is the standard deviation of stratum $k$.
\end{itemize}
Note that $S_k = \sqrt{P_k(1-P_k)}$ is unknown before sampling, where $P_k$ is the political prevalence in stratum $k$. Instead, we use our best estimate, the mean of the range limits of each stratum to calculate the approximate standard deviation expected in each stratum ($P_k = 0.05$ for Stratum 1, $P_k = 0.15$ for Stratum 2 and so on). Since our aim is to accurately estimate prevalence for each subreddit type, we perform separate stratified samplings, choosing two different $W_k$ for each stratum $k$ to match the overall comment proportions over strata (classifier strata profiles) for the two subreddit types.

We modified the Neyman allocation to include a minimum threshold to sample at least 50 comments in each stratum. We added this threshold to reflect the relative uncertainty in our initial estimates of $W_k$ and $S_k$. We had a total budget for rating 10,000 comments. We used $n=2000$ comments from political and $n=8000$ for other subreddits. The rationale behind the uneven breakdown is that there are likely fewer political comments in non-political subreddits, meaning that a similar sized confidence interval for both estimates would result in significantly larger levels of relative uncertainty for prevalence estimates in non-political spaces.  
The fraction of comments that fall into each stratum (the classifier strata profile) are shown in the ``Pop. proportions'' columns of Table~\ref{tab:sampling_estimation} and graphically as a histograms in Figure~\ref{fig:building_calibrators}.
The number of comments selected per stratum are reported in the ``Samples'' columns of Table \ref{tab:sampling_estimation}.

\subsubsection{Results of Stratum-specific prevalence estimation using human judgments}
The prevalence estimates per stratum from labeling comments for the two subreddit types are shown in Table~\ref{tab:sampling_estimation} under ``Political \%'' ($p_{k,\textrm{pol}}$ and $p_{k,\textrm{nonpol}}$). They are also shown graphically in the calibration curves in Figure~\ref{fig:building_calibrators}, where x-axis is the stratum (classifier output) and y-axis is the calibrated prevalence in that stratum.

\begin{figure*}[!ht]
\includegraphics[scale=0.65]{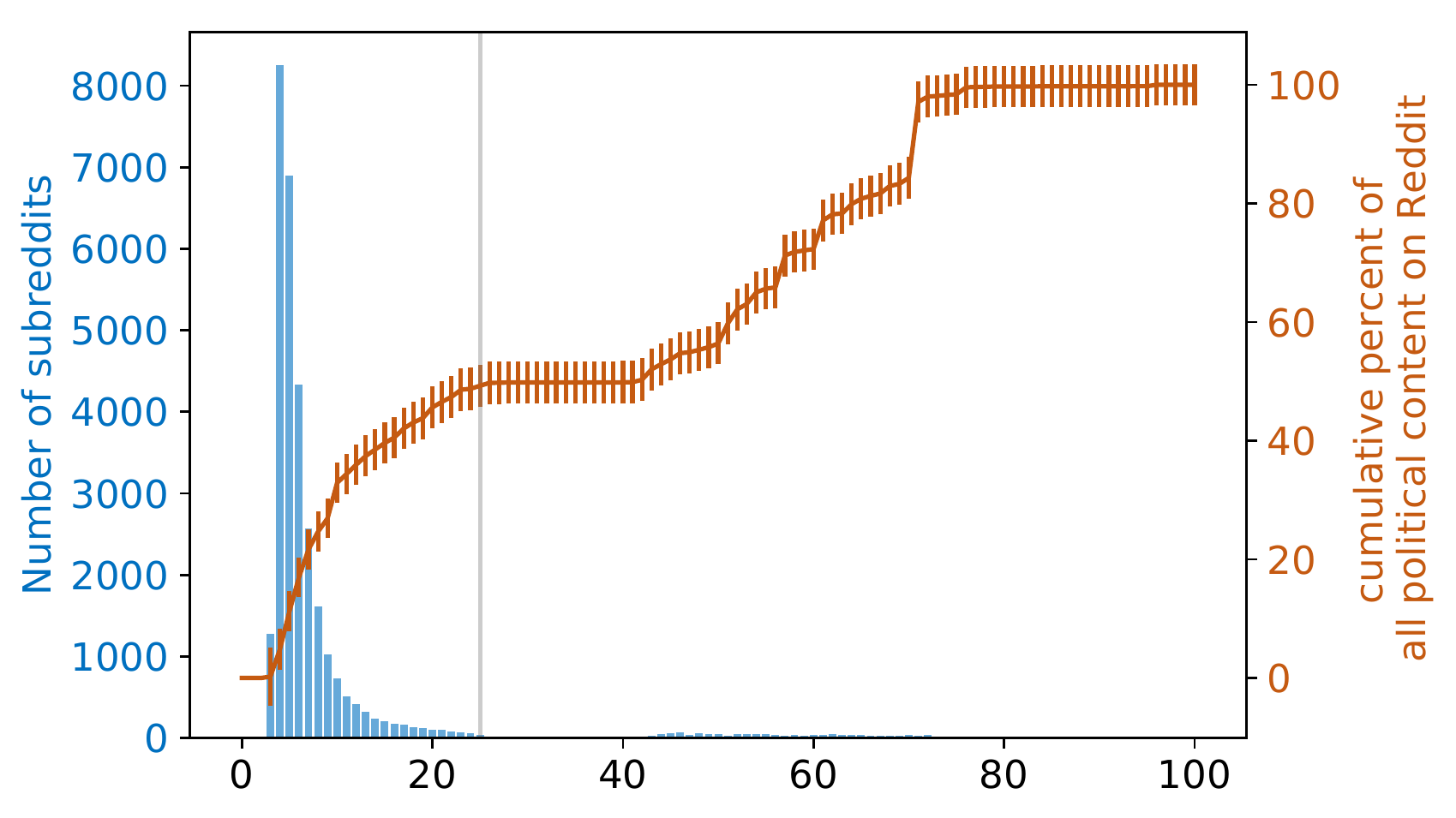} 
\centering
\caption{The line graph shows the cumulative percentage of all political comments posted in subreddits that host political comments less than $x\%$ of the time. The bar graph shows the number of subreddits that host political comments $x\%$ of the time. The grey line marks the 25\% threshold we use to identify subreddits as political or non-political.}\label{RQ1}
\label{fig:subreddit_list}
\end{figure*}

\subsection{Selecting a Calibrator}\label{selectcalib}

Given that the prevalence estimates for the same strata are quite different for the two subreddit types, it is important to determine which set of prevalence estimates to use for each subreddit. Figure~\ref{fig:selecting_calibration} outlines this process. For each subreddit, we obtain the classifier probabilities of all its comments to build its strata profile. If the subreddit strata profile is more similar to the known political strata profile than to the other subreddits strata profile, we use the calibration curve of the known political subreddits, else we use the other calibration curve. We use the Jenson-Shannon divergence (JSD) to make these comparisons between profiles. Lower JSD values imply higher similarity between distributions.

\begin{align*}
\text{diff}_{subr} &= JSD(D_{subr}||D_{pol}) - JSD(D_{subr}||D_{nonpol})\\
 f(subr) &= 
\begin{cases}
    \text{political calibrator},& \text{if } \text{diff}_{subr} \le 0 \\
    \text{non-political calibrator},  & \text{otherwise}
\end{cases}
\end{align*}
where $D_{subr}$, $D_{pol}$ and $D_{nonpol}$ are strata profiles of subreddit $subr$, political and non-political subreddits respectively.

The political and non-political strata profiles shown in Figure \ref{fig:selecting_calibration} correspond to the actual prevalence estimates reported in Table \ref{tab:sampling_estimation}. Comparing the two profiles, we expect that subreddits with strata profiles that are either uniformly distributed or are peaked at around the middle strata will have smaller $\text{diff}_{subr}$ scores, leading to potentially incorrectly assigning calibrators for those subreddits. Reassuringly, we find that less than 2.5\% of all subreddits that we analyze have an absolute $\text{diff}_{subr}$ less than $0.1$, for comparison, the distance between the two strata profiles ($JSD(D_{pol}||D_{nonpol})$) is 0.40. We include the heatmap of subreddits based on $JSD(D_{subr}||D_{pol})$ and $JSD(D_{subr}||D_{nonpol})$ scores in Figure \ref{fig:js_graph} in the Appendix. Further, since in this work it is especially important to not overestimate the prevalence of political content in non-political subreddits, we experimented with a more conservative approach of assigning subreddits to the non-political calibrator (detailed in the Appendix section \ref{ro_calibrator}); we did not find a major difference in the prevalence estimates using this more conservative approach.

\subsection{Corrected ``Classify and Count'' Estimation}\label{estimate}

We quantify the prevalence of political content in each subreddit $subr$ according to the proportion of content in each stratum and the corresponding forecast from the calibration curve. We calculate the estimated prevalence of political content in a subreddit (subr) as:
\begin{displaymath}
p_{\textrm{subr}}= \sum_{k=1}^{K}W_{k,\textrm{subr}} * p_{k,f(subr)}
\end{displaymath}
where $W_{k,\textrm{subr}}$ is the proportion of total comments in stratum $k$ for the subreddit and $p_{k,f(subr)}$ is prevalence estimate of the $k$-th stratum of the calibrator selected by $f(subr)$. 

\subsection{Estimates of Cumulative Counts of Political Comments}\label{cumulative}

We estimate the total prevalence of political content on Reddit as the weighted sum of the prevalence in each subreddit.

\begin{displaymath}
p = \sum \left(\frac{N_{\textrm{subr}}}{N}\right) * p_{\textrm{subr}}
\end{displaymath}
where $N_{\textrm{subr}}$ is the total comments in subreddit $subr$ and $N$ is the total comments across all of the subreddits.

We can estimate the variance of this prevalence estimate by combining the variance estimates across strata. The weight for each stratum is computed from the fraction of comments that the classifier assigns to that stratum. For political subreddits:

\begin{displaymath}
s^2_{\textrm{pol}} =  \sum_{k=1}^{K}{\left(\frac{N_{k,pol}}{N}\right)^2 * s^2_{k,pol}}
\end{displaymath}

where
\begin{displaymath}
N_{k,pol} = \sum_{\textrm{subr} \in \textrm{pol}}N_{k, \textrm{subr}}
\end{displaymath}

$N_{k,pol}$ is the sum of the comments in each stratum $k$ for subreddits similar to the political strata profile. $s^2_{k,pol}$ is the variance estimated for political strata profiles from Table \ref{tab:sampling_estimation}. Similarly, we calculate $s^2_{\textrm{nonpol}}$ for non-political subreddits.

The overall variance is just the sum.
\begin{displaymath}
s^2 = s^2_{\textrm{pol}} + s^2_{\textrm{nonpol}} 
\end{displaymath}

Finally, we define subreddits that are not explicitly political as those that host fewer than some threshold $y$ percentage of political content and calculate the aggregate prevalence and variance of political content in all subreddits that host less than y\% of political content. A higher cutoff of y will, of course, treat more subreddits as non-political and thus yield a higher estimate of the proportion of all political content that is in non-political subreddits. 

\subsection{Prevalence Estimation Results}\label{estimation_results}
In total, we estimate that 12.84\% $\pm 0.45$\%  of all comments on Reddit are political. To study the prevalence of political content in subreddits that are not explicitly political we construct Figure \ref{RQ1}. The blue histogram shows the frequency of subreddits with x-coordinate percentage of political content. Of the 1399 subreddits whose classifier strata profile was closer to the profile for known political subreddits, almost all (99.71\%) were estimated to have 40\% or more political content. 
Of the 29,500 subreddits that were closer to other classifier strata profile, almost all (99.79\%) were estimated to have less than 25\% political content.

Each point on the orange line graph represents the cumulative percent of all political content on Reddit contributed by subreddits that host political comments less than x-coordinate percent of time. We find that 49.26\% $\pm 3.59$\% of all political content on Reddit are from subreddits that host political content less than 25\% of the time. 
Most subreddits on Reddit host very little political content, but cumulatively the non-political subreddits contribute nearly half of all political comments. This could be driven by the few most popular non-political subreddits having far more comments overall than the political subreddits. We examine this possibility by removing the top 10 non-political subreddits (see Table \ref{top10} in the Appendix) that contribute the most political comments. After removing these subreddits, we find that, similar to our original estimates, about 44.82\% $\pm 3.42$\% of all political content on Reddit are from subreddits that are not explicitly political. These results suggest that the large fraction of political content in non-political subreddits is primarily driven by a large number of relatively small subreddits that each host a small percentage of political content. Robustness checks using different human judgment aggregation strategies and calibrator selection approaches yield similar estimates (see Appendix sections \ref{ro_agg_strategy} and \ref{ro_calibrator}).

\section{Quantifying Toxicity of Cross-partisan Political Discussions in Non-political Spaces}

Our main goal in this section is to identify the toxicity levels of cross-partisan discussions on political topics in non-political spaces. 
Our secondary goal is to compare that toxicity to toxicity observed in other settings to better contextualize our findings. To do so, we determine how toxicity on Reddit varies according to the following attributes: (i) political leaning of the users participating in the discussion, (ii) nature of the discussion, and (iii) type of the subreddit where the conversation is taking place. 

For (i), we define a cross-partisan discussion as a left leaning user replying to a right leaning user or vice-versa.
Our analysis is focused on the \textit{reply} comments for each parent-reply discussion pair as the parent comment could be directed at a co-partisan or may not be directed at anyone if it is a top-level comment. 
For (ii), we rely on our calibrated classifier to determine the probability of a reply being political. Finally, for (iii), we classify any subreddit that contains fewer than 25\% political content as not being explicitly political as per Section ~\ref{estimation_results}.

\subsection{Identifying Political Leaning of Users}
To identify political leaning of users, we adopt a simple heuristic similar to ones that have been used in prior Reddit political studies \cite{an2019political,soliman2019characterization}. First, we identify the well known subreddits r/politics, r/Liberal, r/progressive as left-leaning and r/The\_Donald, r/Conservative, r/Republican as right-leaning. Then, we identify a user as left leaning only if all three of the following conditions are satisfied:
\begin{enumerate}
    \item They post more comments in left-leaning subreddits than right-leaning subreddits.
    \item The mean karma points score of their comments in left-leaning subreddits is higher than their mean score in right-leaning subreddits.
    \item Their mean karma score in left-leaning subreddits is greater than 1. 1 is the default score that any comment receives on Reddit. So, a higher than 1 karma score implies that the comment has met net approval by the community.
\end{enumerate}
Similarly, we identify right leaning users. Among users who posted at least once in these subreddits, we have 1,223,229 left leaning and 367,363 right leaning users. We cannot identify  political leanings of other users and do not include them in this analysis.

\subsection{Quantifying Toxicity of Replies}

We use the Perspective toxicity classifier to identify toxicity of a comment. The classifier provides the probability of a comment being toxic, defined as ``a rude, disrespectful, or unreasonable comment that is likely to make people leave a discussion'' \cite{wulczyn2017ex}. The Perspective classifier has been used in prior Reddit studies \cite{mittos2020analyzing,xia2020exploring}. Research evaluating its performance on comments from political communities shows that, on average, its toxicity classification is comparable to a single human judgment of toxicity \cite{rajadesingan2020quick}. We have toxicity classifier probabilities only for comments posted in 2016 and 2017, so we limit this analysis to comments posted in that time interval \footnote{We use the 5th version of the Perspective classifier}.  

We calculate the probability of a reply comment $r$ being toxic and political $TP_r$ as:
\begin{align*}
TP_{r} = toxicity(r) * political(r)
\end{align*}
where $toxicity(r)$ is the toxicity probability given by the Perspective classifier and $political(r)$ is the probability that the comment is political, which is calculated using the calibrated classifier. Similarly, we calculate the probability of the reply comment $r$ being toxic and not political $TNP_r$ as:
\begin{align*}
TNP_r = toxicity(r) *(1-political(r))
\end{align*}

\subsection{Comparing Toxicity Between Discussion Pairs}

Our aim is to compare the mean toxicity levels of cross-partisan political interactions to toxicity levels in other settings. Replies from the same subreddit are clustered to perform semi-pooling. We conduct mixed effects logistic regression using the $lme4$ package \cite{bates2014fitting} modeling the toxicity of replies with a random effect for subreddits. The count of toxic replies is modeled as the number of \textit{successes} and the total replies as the number of Bernoulli \textit{trials} in a binomial distribution.

We estimate the following 3-way interaction model: 
\begin{align*}
  T_{s,polreply,cross} &= Binomial(P(toxicity),\\ 
  & N_{s,polreply,cross})\\
  P(toxicity) &= logit(\alpha_s + \beta_{1}polsub +  \beta_{2}polreply \\
  &+ \beta_{3}cross + \beta_{4}polsub*polreply \\
  &+ \beta_{5}polsub*cross \\
  &+ \beta_{6}polreply*cross \\
  &+ \beta_{7}polsub*polreply*cross)
\end{align*}
where, $polsub$ is an indicator variable for whether the subreddit $s$ is political, $polreply$ denotes whether the reply is political, $cross$ represents whether the reply is directed at an out-partisan. For each subreddit $s$, we identify the total number of replies ($N_{s,polreply,cross}$) for each $(polreply,cross)$ combination and the number of replies in $N_{s,polreply,cross}$ that are toxic ($T_{s,polreply,cross}$). We quantify the total political cross-partisan replies and number of such replies that are toxic in subreddit $s$ as:
\begin{align*}
N_{s,polreply=1,cross=1}&= \sum_{r\in XR_{s}}political(r)\\
T_{s,polreply=1,cross=1}&= \sum_{r\in XR_{s}}TP_r\\
\end{align*}
where $XR_s$ is the set of all cross-partisan replies in subreddit $s$. Similarly, we quantify the non-political co-partisan replies and number of such replies that are toxic in $s$ as:
\begin{align*}
N_{s,polreply=0,cross=0}&= \sum_{r\in CR_{s}}(1-political(r))\\
T_{s,polreply=0,cross=0}&= \sum_{r\in CR_{s}}TNP_r\\
\end{align*}
where $CR_s$ is the set of all copartisan replies in subreddit $s$. Similarly, we calculate $N_{s,polreply,cross}$ and $T_{s,polreply,cross}$ for all other combinations of $(polreply,cross)$.

\begin{table}[!t]
\small
  \centering
  \begin{tabular}{|c|c|r|r|}
    \hline
    \shortstack{Interaction\\type}
    & \shortstack{Conversation\\type} & \shortstack{Political\\subreddits}& \shortstack{Non-political\\subreddits}\\
    \hline
	\multirow{2}{*}{Copartisan} 	& Non-political 	& 13.83M & 57.55M\\
 	 	& Political 	& 21.48M & 6.52M\\
 	 \hline
 	\multirow{2}{*}{Cross-partisan} 	& Non-political 	& 3.72M & 20.51M\\
 	 	& Political 	& 6.03M & 3.17M\\
    \hline
  \end{tabular}
  \caption{Discussion data (in millions) for model estimation.}
  \label{discussion_pair}
\end{table}

Table \ref{discussion_pair} shows the number of replies for each $(polreply,polsub,cross)$ combination used to estimate the binomial model. Upon estimating the model, we are most interested in (i) comparing the average toxicity levels of cross-partisan political discussions in non-political spaces to such discussions in political spaces. We  further (ii) compare the average toxicity levels of political and non-political cross-partisan interactions in non-political spaces, since any large increase in toxicity levels when talking politics has important implications for the health of non-political communities. Finally, we (iii) compare average toxicity levels of cross-partisan and co-partisan interactions as a sanity check. We would expect to see a higher level of toxicity in cross-partisan interactions than in co-partisan ones.

\begin{figure}[!t]
\includegraphics[scale=0.09]{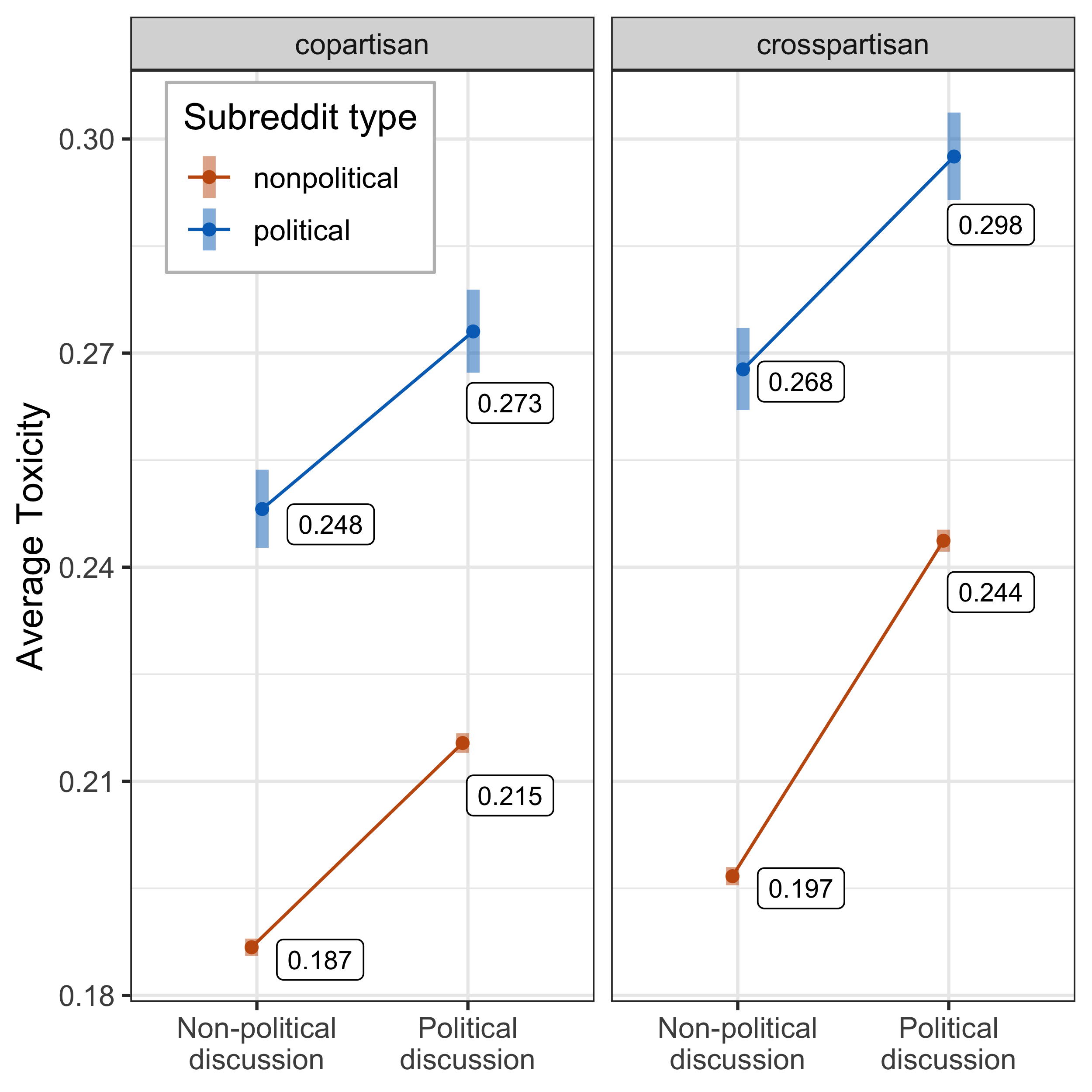} 
\centering
\caption{Interaction plot modeling the toxicity of discussions. All observed pairwise differences are statistically significant at .05 level.}\label{RQ2}
\label{fig:rq2_interactions}
\end{figure}

\subsection{Toxicity Analysis Results}

Since the coefficients in a three-way interaction are hard to interpret, we present our results in the form of interaction plots. In Figure \ref{RQ2}, y-axis is the average toxicity of replies and x-axis is the indicator variable on whether the discussion is political or not. Orange and blue lines represent the toxicity levels in non-political and political subreddits respectively. Left and right subgraphs represent discussions between co-partisans and cross-partisans respectively. 

First, we focus on the plot on the right side of Figure  \ref{RQ2}, considering only cross-partisan replies. Answering our primary research question, we find that cross-partisan replies are significantly less toxic in non-political subreddits (24.4\% toxic) compared to such interactions in political subreddits (29.8\% toxic). 
This holds both for political and non-political content. There is also a main effect of content type, with political discussions being more toxic than non-political ones. 
Finally, there is an interaction effect: the difference in toxicity between political and non-political discussion is significantly larger in non-political subreddits. Still, cross-partisan political replies in non-political subreddits are less toxic than even co-partisan ones, in political subreddits.
A similar pattern holds for copartisan replies. However, co-partisan replies are less toxic than cross-partisan replies in all settings. 

\section{Discussion}

Political discussions on Reddit, much like face-to-face discussions, appear to crop up incidentally in various social settings or communities \cite{conover2018taking}. We found that political discussions, while by definition uncommon in non-political subreddits, are cumulatively abundant. There are many more non-political subreddits than political ones. 
Due to their sheer number, non-political subreddits \textit{cumulatively} host nearly half of all political comments on Reddit.

This result suggests the need to diversify where researchers are looking when they try to understand the nature of political discussion online.  
Importantly, political discussions were not limited to only a few big non-political communities. Rather, we found a large number of small-sized non-political subreddits that had occasional political comments. 
This surely poses an important challenge. Given the sheer scale of content in non-political communities, this necessarily requires building classifiers to accurately identify political content across a wide variety of communities. As is evident from our calibration exercise, commonly used classifiers generally include bias, making this task daunting.

Political conversations in non-political spaces not only add to the volume of total political discourse but also are qualitatively different from conversations in political communities. We find compelling evidence to support the theory that cross-partisan political discussions are indeed much less toxic in non-political spaces than such discussions are in political spaces \cite{wright2015third}. There are multiple potential explanations for this finding. First, political discussions in non-political communities are more likely to be moderated by shared group identity \cite{levendusky2018americans} and social ties \cite{birchall2020trying} instead of partisan identity which may reduce cross-partisan animosity \cite{gaertner2011common}. Second, in general, the toxicity levels observed in non-political communities are much lower and it is likely that these low toxicity norms moderate and temper the tendency to indulge in harsh rhetoric in cross-partisan interactions \cite{iyengar2015fear}.
Regardless of the cause, our findings pose an important caution for researchers: simply aggregating political discussions from political and non-political communities may obfuscate the differences in the types of conversations in these spaces.

There is one important nuance in our toxicity findings. While cross-partisan political discourse is indeed less toxic in non-political spaces, it is significantly more toxic than non-political discourse in the same non-political spaces. Thus, these conversations may have adverse effects on non-political communities. More research is required to understand the consequences of political interactions in these spaces. Further, we observe a larger increase in toxicity levels when talking politics in these spaces than when talking politics in political spaces. We speculate that the norms, rules and the style of moderation in place to foster conducive topic-specific conversations in non-political communities may not be as effective in handling toxicity stemming from cross-partisan political discussions. Further, a political comment in a non-political space can also be seen as a norm violation, leading to aggression from other community members. Alternately, the smaller increase in toxicity levels in political subreddits could indicate a ceiling effect; the toxicity levels of non-political discussions in political subreddits may already be so high that they are near the upper limits of how toxic the discussions can be.

Finally, there are important open questions regarding how these political conversations in atypical spaces fit into the  ``deliberative system'' and how they ought to be studied. The deliberative system consists of both formal spaces such as legislatures and townhalls as well as informal spaces such as social gatherings and online political discussions in social media sites \cite{parkinson2012deliberative}. Recently, deliberation  theorists have highlighted the importance of everyday talk as a web that interconnects these diverse deliberation sites, urging empirical researchers to study discussions wherever they happen \cite{mansbridge1999everyday,conover2018taking}. Future work on how ideas, frames and narratives transition from these spaces to more explicitly political deliberation sites both online and offline will provide important insights on the role and importance of these conversations in non-political spaces.

\section{Limitations and Future Work}

The approach we followed to estimate prevalence is an improvement over a conventional classify and count approach in two important ways, but is still imperfect. The first improvement is that we employ a calibration process to map probabilistic outputs of the classifier into calibrated forecasts of the frequency of political comments. The second improvement is that, rather than assuming that the classifier performs equally well on comments originating in different subreddits, we separately calibrate the classifier on two samples of comments, one from known political subreddits and one from other subreddits. Indeed, we do find that the same classifier score is much more likely to indicate a political comment when the comment comes from a political subreddit, and this dual calibrator approach allows us to appropriately lower estimates of the prevalence of political comments in non-political subreddits.

The approach, however, is still imperfect. First, while creating two calibrators is better than one, there could be more than two types of subreddits, with the classifier having a different error profile for each. Second, our process for selecting the calibrator for each subreddit, by comparing its classifier strata profile to that of the known political subreddits and that of other subreddits, may itself be error prone. We have taken a conservative approach, with more subreddits using the calibrator that yields lower counts than the number of subreddits that are eventually classified as non-political based on their counts. This avoids overestimating the political content in non-political subreddits, but may undercount the political content in political subreddits. 

In the first step, we use a simple n-grams based logistic regression classifier as opposed to using word-embeddings or deep learning techniques. Developing a more accurate classifier generally will improve the effectiveness of the stratified sampling since each stratum is likely to be more homogeneous, leading to smaller confidence intervals for the overall prevalence estimate (see \cite{kumar2018classifier} for a similar argument using a simulation analysis). In our particular case, since we train the classifier on comments from political subreddits rather than on political comments, the gains from using a more accurate classifier are likely tempered by the extent to which comments from political subreddits accurately approximate political comments. Future research examining the gains of using more accurate classifiers in combination with calibrators will refine prevalence estimation techniques. Finally, our robustness checks suggest that the relatively low levels of agreement between raters did not majorly affect our prevalence estimates. Yet, raters disagreeing frequently on what is political indicates scope for improvement in the labeling process, perhaps by providing training examples and exercises.

In the toxicity analysis, we also did not perform a similar calibration process of the Perspective API. While a previous study showed that it was reasonably accurate on content from political subreddits \cite{rajadesingan2020quick}, there was insufficient data provided for calibration, and we do not know whether the error profile of the Perspective API is different on content originating in political vs. non-political subreddits. Numerous other factors, in addition to the type of subreddit, political nature of the comment, and partisanship, can dictate toxicity of responses on Reddit (e.g. toxicity of the parent comment). In our analysis, we used a simplified model to only account for the select attributes of interest to our study. 
Finally, while Reddit is a popular online forum for political discussions, it surely is not the only one. Future work that determines the role non-political communities play in driving political discourse on other platforms can help political communication scholars better identify spaces to pay attention to.

\section{Conclusion}

The subreddits where political comments are uncommon cumulatively produce almost 50\% of all political comments. This is true even when we estimate prevalence based on conservative classifier calibrations. This large cumulative prevalence is not because of the volume of political comments in a few large non-political subreddits; instead, it is driven by a large number of non-political subreddits that host occasional political conversations. Importantly, political comments in non-political spaces seem to be less toxic on average. Thus, scholars looking at the promise and perils of online political deliberation would do well to focus their attention on political discussions that occur in venues that are not primarily organized to encourage political discussion.

\section*{Acknowledgements}
This material is based upon work supported by the National Science Foundation under Grant No. IIS-1717688. Ashwin Rajadesingan is supported by a Facebook Fellowship. We thank the ARC-TS team at Michigan for Cavium-Thunderx Hadoop cluster support. We also thank the anonymous reviewers for their invaluable feedback on this work.

\section*{Appendix}
\renewcommand{\thesubsection}{\Alph{subsection}}
\subsection{Robustness of Prevalence Estimates Varying Label Aggregation Strategy}\label{ro_agg_strategy}
We estimate prevalence with two other aggregating strategies: (i) `any one' strategy: comment is political if at least one rater labels it as political. (ii) `all three' strategy: comment is political if all three raters label it as political. We estimate that 54.56\% $\pm 3.28\%$ and 42.28\% $\pm 3.95\%$ of the overall political content are from subreddits that are not explicitly political, based on the `any one' and `all three' strategies respectively.

\subsection{Robustness of Prevalence Estimates Varying Political Subreddit Identification Strategy}\label{ro_calibrator}
To ensure robustness of our prevalence estimates, we employ another identification strategy. We relax the $\text{diff}_{subr}$ criterion such that subreddits near the decision boundary will be calibrated using the political calibration curve. With this strategy, the prevalence estimates in political subreddits are expected to be higher.:

\begin{displaymath}
 f(s) = 
\begin{cases}
    \text{political calibration curve},& \text{if } \text{diff}_{subr}\leq 0.1\\
          \text{non-political calibration curve},  & \text{otherwise}
\end{cases}
\end{displaymath}
From this, we find that 43.29 $\pm 3.40\%$ of overall political content are from subreddits that are not explicitly political. 

\begin{figure}[!ht]
\includegraphics[scale=0.77]{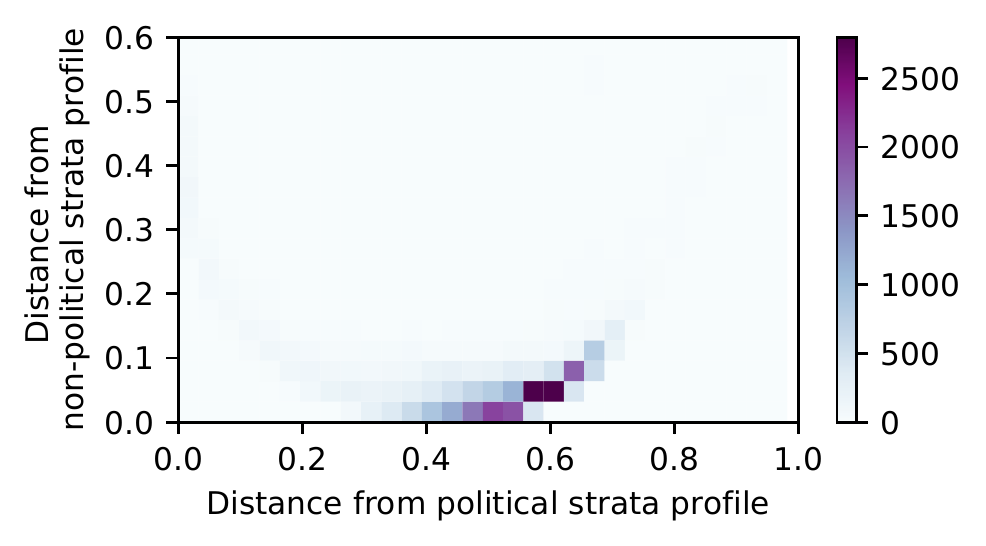} 
\centering
\caption{Density of subreddits by $JSD(D_{subr}||D_{pol})$ and $JSD(D_{subr}||D_{nonpol})$ scores. Few subreddits are equidistant from both strata profiles; most are close to the non-political strata profile. }\label{fig:js_graph}
\end{figure}

\begin{table}[!h] 
\small

  \centering
  \begin{tabular}{|c|r|r|}
    \hline
 Subreddit & \shortstack{Political\\ percent} & \shortstack{Political\\ comments} \\
    \hline
AskReddit & 9.97 & 15,151,228 \\
pics & 19.15 & 2,979,427 \\
todayilearned & 16.57 & 2,537,532 \\
unpopularopinion & 22.49 & 1,995,731 \\
videos & 12.01 & 1,632,210 \\
funny & 10.03 & 1,614,550 \\
nba & 5.73 & 1,481,271 \\
nfl & 6.32 & 1,373,542 \\
soccer & 6.71 & 1,223,417 \\
AdviceAnimals & 20.94 & 1,080,184 \\

    \hline
  \end{tabular}
  \caption{Non political subreddits that contain the most political comments }\label{top10}
\end{table}

\bibliography{main}

\end{document}